\documentclass[journal,12pt,onecolumn,showkeys]{revtex4-1}
\usepackage{graphicx}
\usepackage{amsmath}
\usepackage{amssymb}
\usepackage{dcolumn}
\usepackage{latexsym}
\usepackage{rotating}
\usepackage[usenames,dvipsnames]{xcolor}
\usepackage{float}
\usepackage{epsfig}
\usepackage{psfrag}
\usepackage{natbib}
\usepackage{bm}
\usepackage{eucal}
\usepackage{braket}
\usepackage{enumerate}
\usepackage{longtable}
\usepackage{subfigure}
\usepackage{bm}
\usepackage{hyperref}
\usepackage{amsfonts}
\usepackage{dcolumn}% Align table columns on decimal point
\usepackage{bm}
\usepackage{subfigure}
\newcommand{\be}{\begin{equation}}
\newcommand{\ee}{\end{equation}}
\newcommand{\bn}{\begin{eqnarray}}
\newcommand{\en}{\end{eqnarray}}

\usepackage{color} % use colored text in latex

\usepackage{hyperref}
\begin{document}

\author{S. Koley$^{1}$}\email{sudiptakoley20@gmail.com}
%\author{Saurabh Basu$^{2}$}

\title{Intercalation in 2H-TaSe$_2$ for modulation of electronic properties and electrochemical energy storage}
\affiliation{$^{1}$ Department of Physics, Amity Institute of Applied Sciences, Amity University,  
Kolkata, 700135, India}
%\affiliation{$^{2}$ Department of Physics, Indian Institute of Technology Guwahati, Assam, 781039, India}
\begin{abstract}
\noindent Two-dimensional transition metal dichalcogenides (TMDs) exhibit 
an extensive variety of novel electronic properties, such as charge density 
	wave quantum spin Hall phenomena, superconductivity, and Dirac and Weyl 
semi-metallic properties. The diverse properties of TMDs suggest that 
structural transformation can be employed to switch between different electronic
	 properties. 
Intercalation and zero valence doping of molecules and atoms into 
the van der Waals gap of TMDs have emerged as effective approaches to modify the 
charge order states of the material. This eventually leads to phase transition 
or the formation of different phases, thus expanding the electronic, 
thermoelectric and optical applications of these materials.  
In this study, electronic and electrochemical energy storage
 properties of such an intercalated TMD, namely, 2H-TaSe$_2$ via intercalation 
	of lithium (Li), sodium (Na) and potassium (K) have been investigated. 
	The intercalation of these ions into the dichalcogenide resulted in a modified band structure and novel structural effects, leading to the emergence of a 1 eV band gap. Possibility of electrochemical energy storage application is 
	also explored in this study. Furthermore, the importance of multi 
	orbital electron-electron correlations in intercalated TaSe$_2$ is 
	also investigated via dynamical-mean-field theory with local density 
	approximation.

\end{abstract}
%\pacs{
%71.10.Hf,
%Non Fermi Liquid
%}
\maketitle

\section{INTRODUCTION}
Energy storage particularly in the form of rechargeable batteries, poses a 
critical limitation that impedes the progress of various technologies. 
From biomedical applications and portable electronic devices to emerging 
electric vehicles and 
renewable energy storage, the primary hindrance to advancing these fields is 
the lack of an efficient medium for energy storage. 
Batteries already play a significant role in modern life already. 
Despite substantial research in rechargeable battery technology, the problem 
of short battery life persists in many of our daily appliances.
Over the past few decades two-dimensional materials with a diverse range of 
physical properties have gernered widespread attention from energy storage 
researchers\cite{novo,mak,Wang,shin}. The use of aqueous 
multivalent metal-ion batteries (AMMIB) has become a well-established
approach to address the rapidly increasing demand for high performance and
cost effective storage devices.

Transition metal dichalcogenides have 
remarkable potential in the development of innovative electronic devices owing 
to their distinctive
geometries and efficient charge transfer properties\cite{manzeli}. The monolayers of these materials 
 are bound by van der Waals interactions. Ample electronic and electron phonon 
 interactions in TMDs lead to strong correlation. Additionally, charge 
 density wave (CDW) and superconducting ground states are found frequently in 
 their phase diagram.
 2D materials are also one of the most fascinating topics of research due to 
 the complex nature and significant quantum fluctuations of the collective 
 electronic states. The common chemical 
 formula of TMD is MX$_2$, where M represents a transition metal and X denotes 
 a chalcogen. 
 The layered structure of TMDs leads to excellent host for intercalation due to
 the weak interlayer van der Waals force.

Intercalation is a familiar occurance in layered chalcogenides 
and is essential for variety of applications, such as, supercapacitors, 
battery electrodes, and 
solid lubricants\cite{wang}. The weak interlayer force enables insertion of ion, thereby promoting the multivalent ion intercalation in AMMIBs.    
Recent studies have attested that intercalated chalcogenides
can induce superconductivity in the parent material\cite{srxbi,biag} in addition to serving as an effective medium for electrochemical energy storage.  
The intercalation process is easy and reversible without altering the structure
 of parent materials. Potential reversibility of this process facilitates in 
 charging and 
 discharging processes in intercalated battery structure\cite{sengupta}. 
 The concentration of intercalation varies with electrochemical potential or 
 temperature leading to changes in physical properties. 
 Aditionally, this process provides a high 
 doping concentration in 2D chalcogenides and can modify CDW ordering
 of a material, resulting in the appearance of another ordered (superconducting) state in the  
 phase diagram of TMD materials\cite{xing}. Theoretical simulations 
 predict that extra carriers will influence electronic dispersion and phonon band structure in TMDs, resulting in either suppressed CDW state\cite{Jliu} or 
 modified CDW transition temperature. 

Transition metal dichalcogenides (TMDs) have garnered extensive attention for 
their potential applications in energy storage devices because of their 
favorable chemical and physical properties as well as their wide interlayer 
distance\cite{stephenson,Yu,Chen}. 
Recent theoretical studies suggested that MoS$_2$, MoSe$_2$, WS$_2$ and 
their heterostructures possess promising energy storage capabilities\cite{wu,yun}. These dichalcogenides are observed to retain much higher
 theoretical capacity than traditional batteries. Despite significant 
 advancement in TMD-based cathode material development, further research is 
 ongoing to devise innovative TMD cathode materials that can surmount low 
 conductivity issues in intercalated chalcogenides. To the best of author’s 
 knowledge, no studies have yet reported theoretical capacity of lithium, 
 magnesium and calcium intercalated 2H-TaSe$_2$. 2H-TaSe$_2$ is a material 
 that possess zero bandgap in its density wave ordered state, thereby indicating higher conductivity in its intercalated phases. Consequently, to verify its 
 theoretical storage capacity and conductivity it is imperative to determine 
 the electronic properties and binding energy of the 
 intercalated 2H-TaSe$_2$. 

 In this study, intercalation in 2H-TaSe$_2$ have been investigated and the unique properties  
 of the resulting intercalated dichalcogenide as well as its potential
 applications in energy 
 storage and optoelectronics have been analysed. 2H-TaSe$_2$ features a
 hexagonal structure with Se-Ta-Se tr-ilayer stack and is classified as a 
 metallic TMD exhibiting CDW 
 transition temperature at 122K and very low superconducting transition 
 temperature of 0.15K. Also the interlayer distance of 0.4 nm of 2H-TaSe$_2$ 
 renders it as a suitable host for intercalation.

\section{Methods}
\noindent In this work, intercalation in 2H-TaSe$_2$ is investigated through the utilization of density functional theory (DFT) calculations  
based on the 
Perdew-Burke-Ernzerhof (PBE) method. 
The generalized gradient approximation (GGA) is employed 
for the exchange-correlation functional, while the full potential 
linearized augmented plane wave (FP-LAPW) method, as 
implemented in WIEN2K package\cite{w2k}, is used for the calculations. 
A kinetic energy cutoff of 20 eV is 
applied for the plane waves to compute the electronic wave functions. 
The integration in momentum space is conducted through a 
$12\times12\times1$ Monkhorst-Pack k-point mesh centered around $\Gamma$-point. 
Furthermore, all the lattice constants and atomic coordinates are fully relaxed 
until the forces 
were smaller than 0.001eV/$\AA$. 
%The force-constant method and the 
%PHONOPY package\cite{phonopy} were used to get phonon spectra which shows real 
%frequency dispersion and hence proves the 
%stability of the doped material. 
%The dynamical matrices were calculated using 
%the finite differences method for doped borophene in large supercell structure.
Self consistent field (SCF) calculations are implemented to calculate the
total energy of intercalated and pristine compounds. In addition,
the optical conductivity, and imaginary and real parts of dielectric 
tensor were computed within the WIEN2K code. The 
interstitial wave function is expanded in terms of plane waves 
with a cut-off parameter of R$_{MT}\times$ $K_{max}$=8.5, where R$_{MT}$ and 
K$_{max}$ represent the smallest
atomic sphere radius and the largest k vector in
the plane-wave expansion. The optical terms are evaluated using a 
dense k-point mesh of $15\times15\times1$ $\Gamma$-centered Monkhorst-Pack with 
a fixed Lorentzian broadening of 0.05 eV. 
\section{Results} 
The electronic properties of the dichalcogenide will exhibit modifications due 
to the presence of 
intercalants. These modifications can be attributed to two mechanisms, 
(i) lattice modulation and (ii) transfer of charge from suitable intercalants
to host atoms. The transfer of charge between the host atom and intercalant 
has the potential to alter both the Fermi surface and conductivity of the 
parent material. Consequently, intercalation induced changes in the Fermi 
surface can lead to transition between states such as metal-insulator, 
metal-semiconductor and even superconductivity.
This study focuses on the changes observed in non-interacting structure of 
2H-TaSe$_2$ resulting from
intercalation of three different intercalants: lithium, sodium, 
and potassium.
FIG.1 illustrates crystal structure of lithium intercalated sample. The 
occupancy fractions for Li were determined by considering the total 
concentration of lithium and taking into account the 
multiplicity of the positions. The same constraint was applied to the 
intercalation of Na and K in 2H-TaSe$_2$.
In FIG. 2(a-c) depict the band structures of intercalated TaSe$_2$ obtained 
through LDA calculations. 
While the original band structure displays a reasonable crossing of Ta-d 
orbital at the Fermi level, all doping cases exhibit an opening of 
energy gap initiating from the Fermi level. Intercalation induced bandgap, 
measured at 1 eV,  subsequently introduces alterations in electronic, optical and energy storage properties of the material.

The density of states depicted in FIG.3(a-c) serves as further evidence for the existence of a gap of approximately 1 eV near the Fermi energy. 
The graphical representations of energy band structure and density of states for all 
doping cases exhibit similar qualitative features. However, quantitative 
differences will 
produce different optoelectronic and electrochemical energy storage properties.
\begin{figure}
\epsfig{file=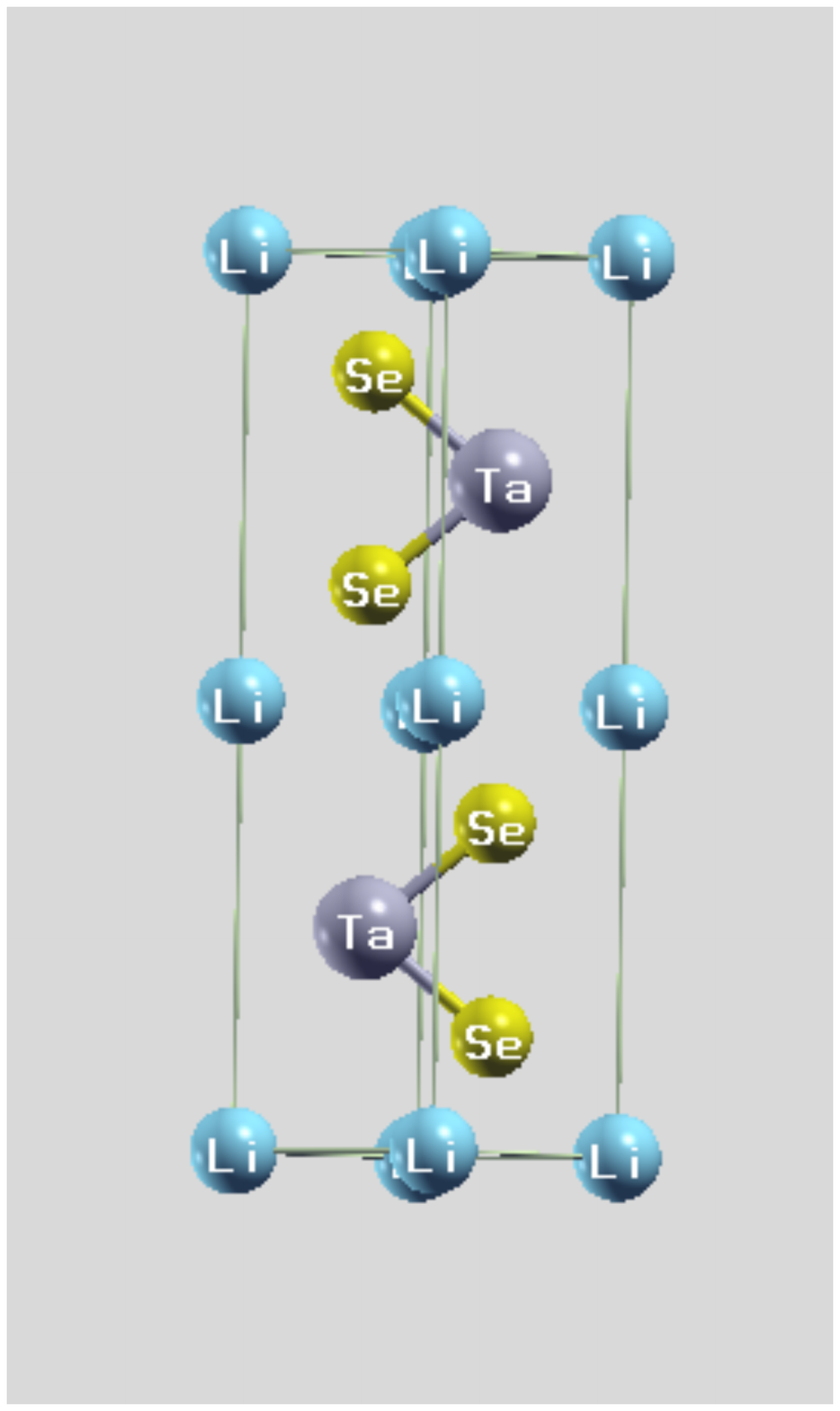,trim=0.0in 0.5in 0.5in 0.0in,clip=false, width=60mm}
\caption{(Color Online) Crystal structure of lithium intercalated 2H-TaSe$_2$. 
	The atoms are marked and colored for reference.
}
\label{fig1}
\end{figure}
\begin{figure*}
\epsfig{file=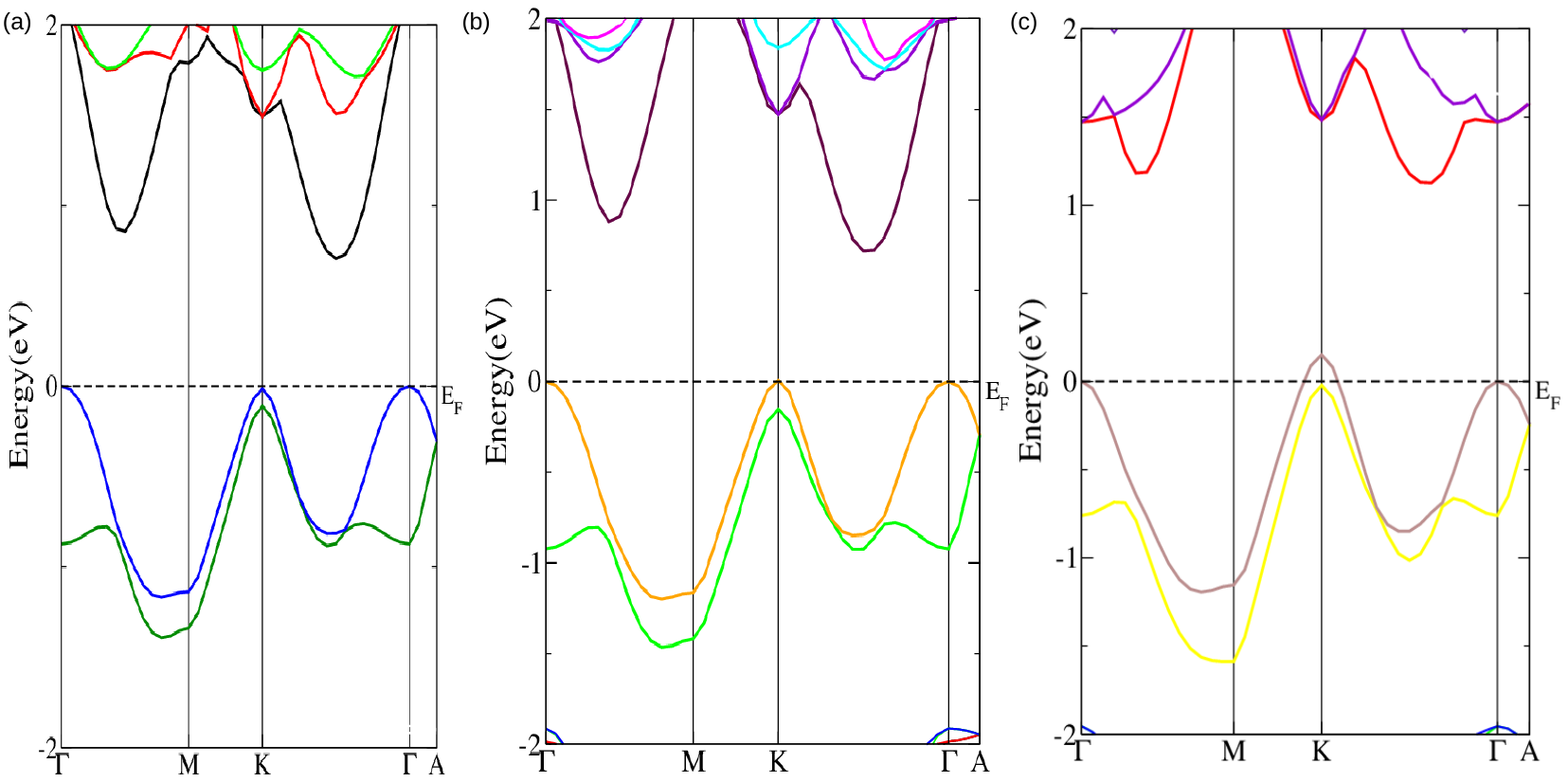,trim=0.8in 2.0in 1.8in 0.3in,clip=false, width=160mm}
\caption{(Color Online) Electronic band structure of intercalated dichalcogenide(2H-TaSe$_2$) with three different intercalation (a) lithium, (b)sodium and (c)
	potassium. 1 eV gap in the band structure is a consistent phenomena in all the electronic structure.
}
\label{fig2}
\end{figure*}

\begin{figure*}
\epsfig{file=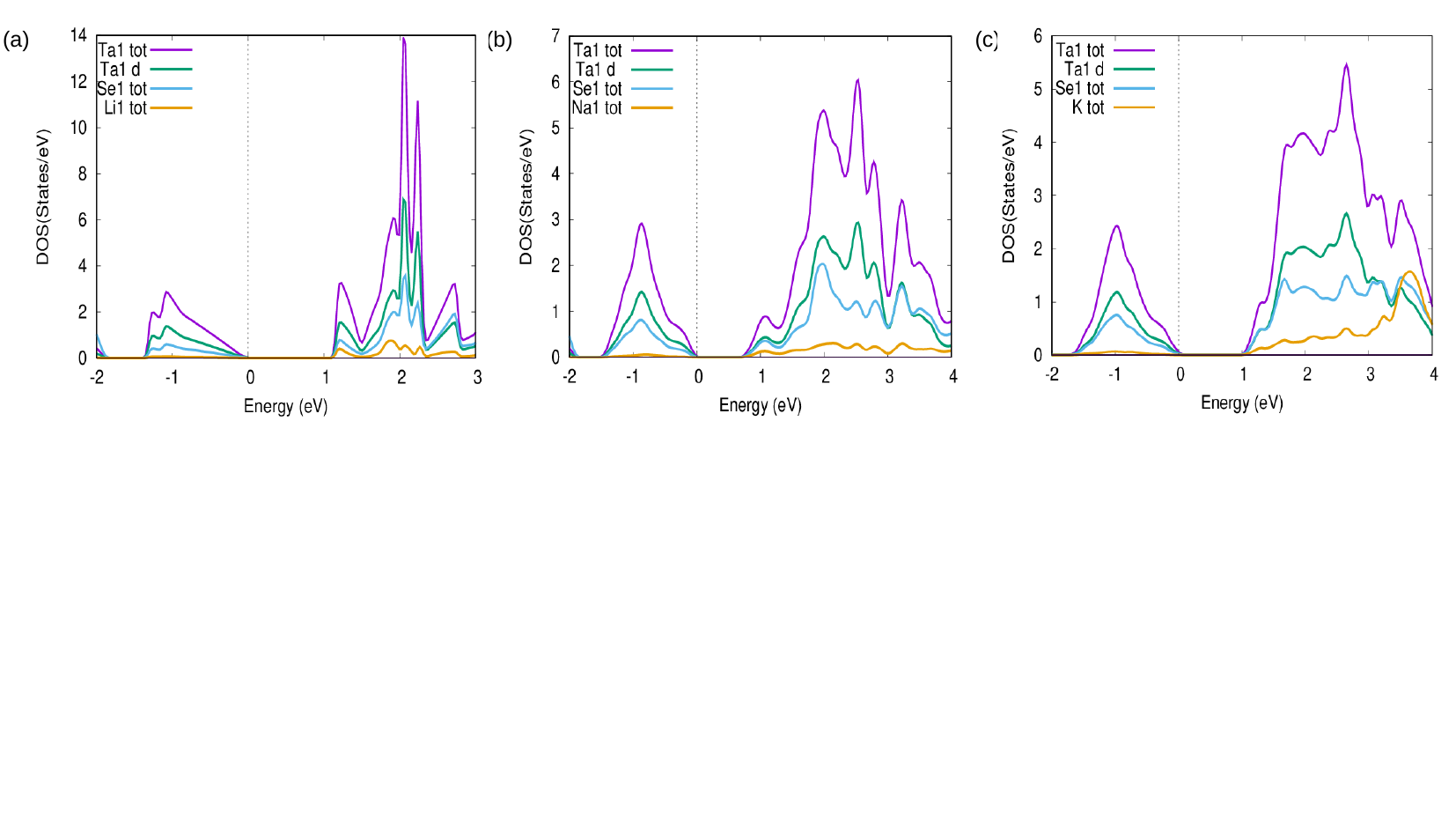,trim=0.0in 3.0in 0.0in 0.0in,clip=false, width=180mm}
\caption{(Color Online) Density of states for intercalated 2H-TaSe$_2$ for 
	(a) lithium, (b) sodium and (c) potassium. Density of states also reveal a band gap of 1 eV and dominant band as Ta-d band.}
\label{fig3}
\end{figure*}
\begin{figure*}
\epsfig{file=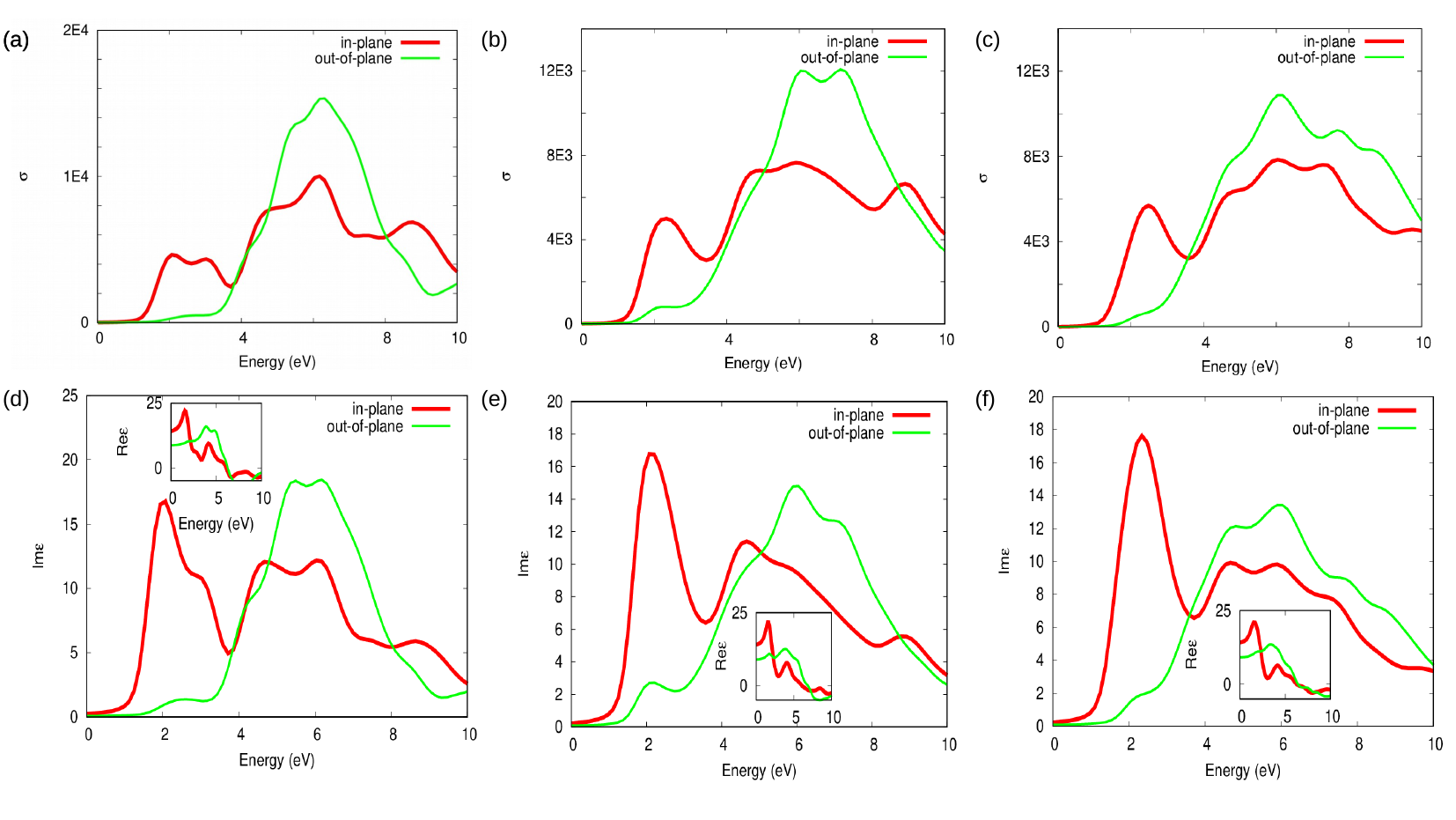,trim=0.0in 0.0in 0.0in 0.0in,clip=false, width=170mm}
\caption{(Color Online) In plane and out-of-plane optical conductivity of 
	(a)Li, (b)Na, and (c)K intercalated 2H-TaSe$_2$. Imaginary part of dielectri constant of the intercalated 2H-TaSe$_2$ is shown in ((d), (e) and (f) respectively), insets show real part of the dielectric constant of the same. }
\label{fig4}
\end{figure*}

Optical conductivity ($\sigma(\omega)$) arises from the motion of carriers with 
oppositely charges induced by electromagnetic wave and, in turn, reflects the 
absorption coefficient and refractive index. In this study, the optical 
spectra are presented for a wide range of energy [0-10 eV] at room temperature.
The optical conductivity plots [FIG.4a-4c] for doped 2H-TaSe$_2$ confirm a 
shifted Drude peak due to the emergence of 1 eV band-gap. Notably, the new 
position of the first peak in the optical conductivity falls within the visible 
region, thereby increasing the potential applications of this novel 
semiconductor material in optoelectronics. The in-plane and out-of-plane 
optical conductivities exhibit significant asymmetry. 
This observation aligns with earlier findings for the parent material as 
well\cite{dordevic}. 
Specifically, the in-plane conductivity displays a small peak in the visible 
region while the out-of-plane conductivity exhibit a prominent peak in the 
higher energy range (4-8 eV). The optical conductivity plots effectively 
demonstrate the intercalation induced changes in the electronic structure 
which have been well-established.

 The frequency-dependent dielectric functions which account for both the 
 interband and  
 intraband contribution can be expressed as $\epsilon(\omega)=\epsilon_1(\omega)+i\epsilon_2(\omega)$ 
 where $\epsilon_1(\omega)$ and $\epsilon_2(\omega)$ represent the real and 
 imaginary 
 parts of the dielectric function, respectively. The imaginary part, $\epsilon_2(\omega)$ 
 is computed through density functional theory band structure calculations. 
 The real part of the dielectric function is obtained from the imaginary part  
 using Kramers-Kroning relation.
 The energy dependencies of these functions are illustrated
 in FIG.4d-4f for the intercalated 2H-TaSe$_2$ with lithium(Li), sodium (Na)
 and potassium (K).
 The plots of the in-plane and out of plane dielectric function demonstrate 
 sizable anisotropy in agreement with optical conductivity data. Notably, 
 Drude peak is absent from zero frequency of imaginary dielectric function. 
 The real part of dielectric function exhibits a high static dielectric 
 constant. This 
 observation indicates weak electron-hole interactions due to strong screening effect.
\section{Electrochemical Energy Storage}
In two dimensional dichalcogenides composed of transition metals, a metallic 
layer is sandwiched between the chalcogen layers. Li, Na and K ions are 
intercalated in between two dichalcogenide layers. The structure of 2H-TaSe$_2$ 
belongs to space group of 194, leading to the selection of adsorption sites 
with fixed space group symmetry. As shown in FIG.1 the intercalants are placed 
in the site with 3m symmetry (0,0,0) and (0,0,1/2) position. After exploring 
other potential sites, this particular site is chosen following comprehensive 
structural relaxation as well as energy minimization as detailed in the methods section. 
The binding energy of intercalant atoms is defined as:
$$E_b=(E_{compound}-nE_{intercalant}-E_{TaSe_2})/n$$
E$_{compound}$ is the full energy of the intercalated TaSe$_2$, in which n is  
no of intercalants added on pristine 2H-TaSe$_2$. E$_{intercalant}$ is the 
energy of an isolated Li, Na and K ions in a vacuum. E$_{TaSe_2}$ is the energy 
of an solitary 2H-TaSe$_2$ layer. 
The more negative the binding energy, the more favorable the structure. In the 
case of Li, Na and K with x=0.2, the calculated binding energies are -1.61 eV, 
-1.27 eV and -0.882 eV respectively. This points to strong attractive 
interactions between the intercalants and the parent structure. In order to 
assess the potential for electro-chemical energy-storage applications, 
the theoretical 
capacity is determined using the formula $$C=cnF/M_{TaSe_2}$$. In the above 
 equation $c$ 
 is the number of adsorbed ions on a single unit 2H-TaSe$_2$, 
 $n$ is the valence state of ions, $F$ is the Faraday constant (26801 mA.h.mol$^-1$), and $M_{TaSe_2}$ is the molar weight of 2H-TaSe$_2$. 
 In this case, both $c$ 
 and $n$, are assumed to be 1 for the intercalations. As a result, the 
 adsorption capacity for 2H-TaSe$_2$ is determined to be 79.1. 
 The electrical conductivity of the intercalated 2H-TaSe$_2$ is obtained from
 Boltztrap code\cite{boltz}. The electrical conductivity 
 of intercalated 2H-TaSe$_2$ demonstrates a consistent upward trend 
 with an increase in the atomic no of intercalants. Based on the considerations of 
 binding 
 energy and theoretical storage capacity, it can be concluded that Li 
 intercalation is more favorable than 
 Na and K intercalation.

\begin{figure*}
\epsfig{file=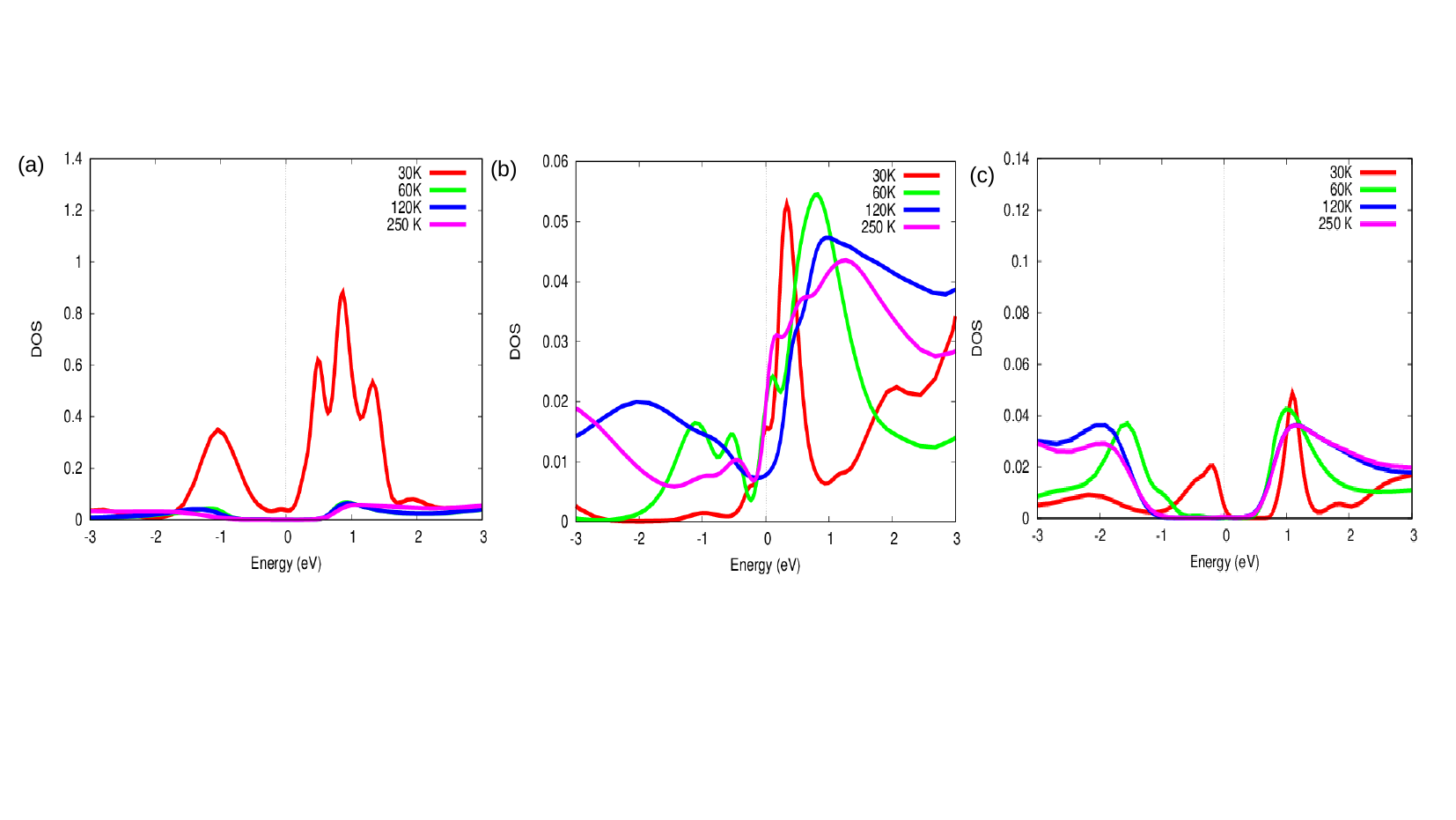,trim=0.0in 0.0in 0.0in 1.0in,clip=false, width=180mm}
\caption{(Color Online) Dynamical mean field theory density of states for four 
	different temperatures (30K, 60K, 120K and 250 K) in three intercalated 
	structures ((a)Li, (b) Na and (c) K intercalated 2H-TaSe$_2$).}
\label{fig5}
\end{figure*}
\begin{figure*}
\epsfig{file=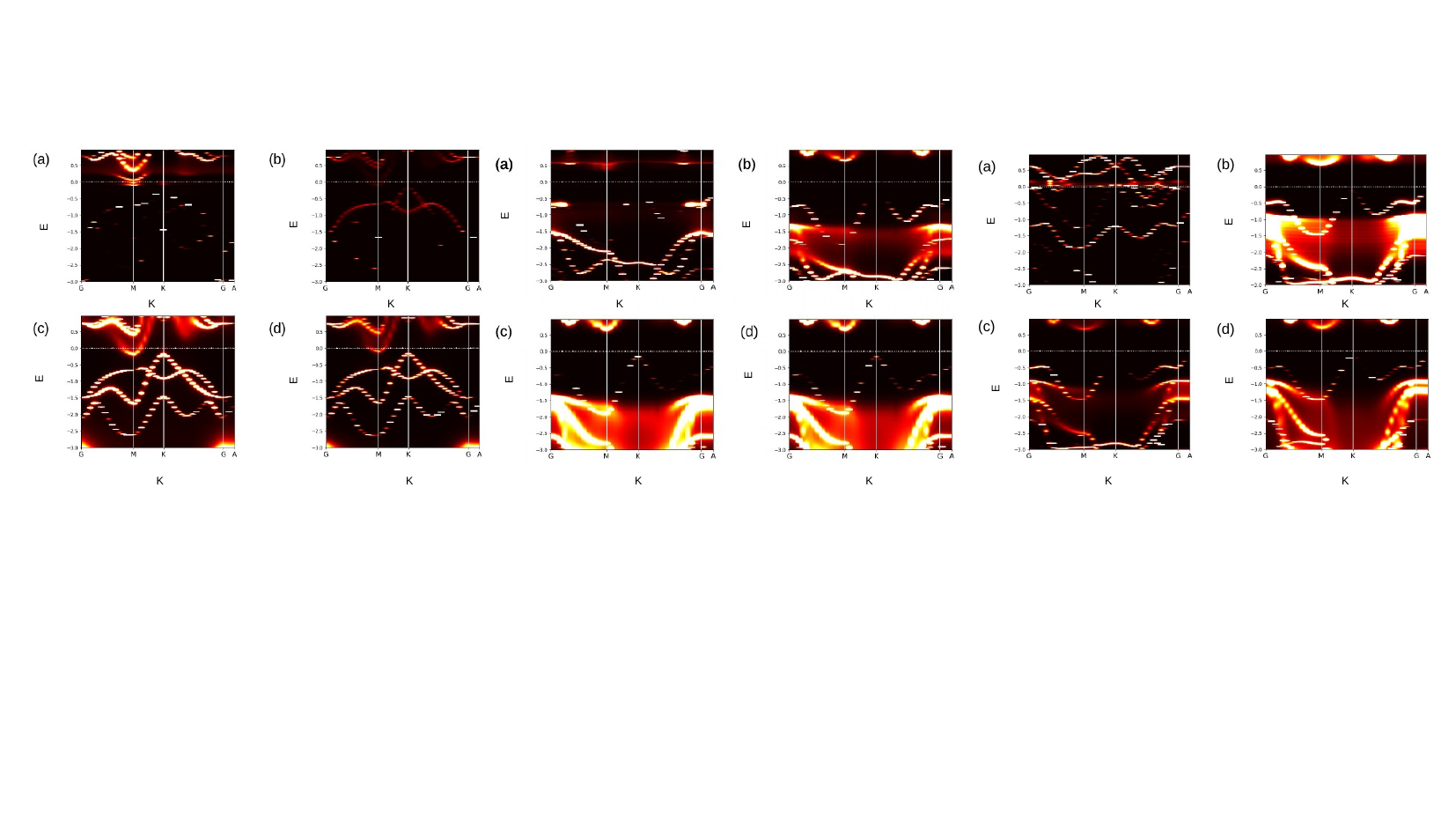,trim=0.0in 2.0in 0.0in 0.0in,clip=false, width=180mm}
\caption{(Color Online) Angle Resolved photoemission spectroscopy energy-momentum plots for Li((a)-(d)) intercalated, Na ((e)-(h)) and K((i)-(l)) intercalated 2H-TaSe$_2$. Each intercalation is studied for four temperatures 30K,((a), (e) and (i)), 60K((b),(f) and (j)), 120 K((c), (g), and (k)) and 250 K ((d), (h) and (l)).  }
\label{fig6}
\end{figure*}

\section{Dynamical Mean Field Theory Calculation}
As dichalcogenide materials exhibit strong correlation, a fully 
charge-self-consistent dynamical mean field theory (DMFT) is used to calculate 
electronic properties. The combination of DFT and DMFT is a well-established 
method to explain important physical properties\cite{at1,at2,at3,at4,at5} of 
these materials. 
In this study, DMFT is implemented using the EDMFTF package\cite{haule1,haule2} which 
integrates the DFT formalism based on WIEN2K . The package effectively handles 
the issue of exact double-counting. The intra- and inter-orbital Coulomb 
interaction are treated accurately and self-consistent determination of the 
imaginary frequency Green's function is performed. The extended Hubbard model 
incorporating both intra and interorbital Coulomb interaction is employed with 
a realistic range of values for these interactions. The total Hamiltonian is 
solved using the DMFT method from low to room temperature. The correlated 
Ta-d orbitals are treated dynamically within DMFT using an orbital 
projection-embedded scheme facilitated by the package while the p orbitals are 
treated at the DFT level. The continuous time quantum Monte Carlo (CT-QMC) in 
the hybridization expansion method is used as impurity solver of the DMFT code. 
Parameters such as temperature, Coulomb interactions U, and U$^\prime$ are 
varied to 
obtain temperature dependent spectral function of the intercalated system. 
Convergence precision of 0.0001 is maintained with respect to chemical 
potential, self-energy and charge density in these calculations for the 
intercalated structure. Finally, the maximum entropy method\cite{maxent} is utilized 
for the analytical continuation from the imaginary axis to real frequencies. 
The resulting real frequency spectral function is then presented in FIG.5 for 
the intercalated structures. FIG.5a portrays the DMFT spectral function for Li 
intercalated 2H-TaSe$_2$. The presence of band gap, except at 30 K, 
is evident from the plot. FIG.5b and FIG.5c depict the energy and temperature 
dependent spectral functions of Na and K intercalated 2H-TaSe$_2$, respectively.
While Na intercalation displays metallic behaviour across all temperatures, K 
intercalation shows a gap in the Fermi energy except low temperature. The DMFT 
spectral functions records the redistribution of spectral density due to 
dynamical correlations, which is missed by static mean-field theory. Static 
mean field theory fails to account for incoherent states and is therefore 
inadequate for describing strong correlation.
Moreover, angle resolved photoemission spectroscopy (ARPES) band-structure maps 
are presented in FIG.6. For lithium doping (FIG.6a-6d) ARPES spectra exhibit 
striking difference at a low temperature (30 K) compared to the other 
temperatures. At very low temperature, the conduction band and valence band 
are overlapping indirectly with each other, but they separate and open a gap 
at the Fermi energy at all other temperatures. Further analysis of the ARPES 
map reveals that the valence band remains pinned at K point, consistent with 
the LDA band structure. However, the valence band shifts downward at $\Gamma$ 
point after dynamical mean field calculation. Regarding sodium doping 
(FIG.6e-6h), the conduction band is observed to cross the Fermi level at the $M$
point across the entire temperature range, while the valence band remains 
closer to Fermi level. This configuration allows for electron-hole interaction 
with a small value of hybridization constant. As a result, even the parent 
2H-TaSe$_2$ system favours electron hole interaction and formation of exciton. 
The ARPES maps for K doping are plotted in FIG.6i-6l. The putative valence band 
is just close to the Fermi level while the conduction band is above the valence 
band at all temperatures. Unlike lithium and sodium doping, potassium doping 
induces a gap in the Fermi energy along $\Gamma - M - K - \Gamma$ direction. 
Overall, the ARPES data provides insight into the ordering in intercalated 
2H-TaSe$_2$.
The findings of the DFT and DFT plus DMFT calculation are summarized in 
conclusion. The intercalated 2H-TaSe$_2$ exhibit a significantly different 
state 
compared to its pristine form. The ground state of the intercalated 
dichalcogenide is characterized by strong correlations and 1 eV bandgap. This 
material holds promise as a new semiconductor material in optoelectronics, 
given its suitable bandgap for the visible part of the optical spectrum. 
Furthermore, a substantial static dielectric constant is observed. The study of 
Li, Na and K intercalation also aims to explore alternative energy sources. 
The energy storage calculations reveal an adsorption capacity of 79.1 for the 
intercalated dichalcogenide with ions of single valence state. This confirms 
the prospect of using these materials as alternate electrochemical energy 
storage material. Henceforth this theoretical research highlights the 
prospective application for fine tuning the properties of 
2H-transition metal dichalcogenide.
\section{Data availability}

The data will be available on request.

\section{Acknowledgement}
S. Koley acknowledges department of science and technology women scientist grant
SR/WOS-A/PM-80/2016(G) for WIEN2K software.

\end{document}